\documentclass[11pt,twoside]{article}
\usepackage{asp2010}

\resetcounters

\begin{document}

\title{FIR extended emission from cold gas and dust in Blue Compact Dwarf Galaxies: the anomalous cases of POX~186 and UM~461\footnotemark}
\author{Vanessa~Doublier$^1$,  Aur{\'e}lie~R{\'e}my-Ruyer$^2$ and SAG2~consortium
\affil{$^1$Max Planck for Extraterrestrische Physik, Giessenbachstr. 1, 85748 Garching, Germany}
\affil{$^2$Laboratoire AIM, CEA/IRFU/Service dÕAstrophysique Universit{\'e} Paris Diderot Bat. 709, 91191 Gif-sur-Yvette, France}}

\allowtitlefootnote

\titlefootnote{{\it Herschel} is an ESA space observatory with science instruments provided by European-led Principal Investigator consortia and with important participation from NASA.}

\begin{abstract}
FIR observation of BCD galaxies with Herschel has revealed a wealth of new insights in these objects which are thought to resemble high-redshift forming galaxies. Dust and cold gas showed to be colder, in more or less quantities than expected and of uncertain origin. However, not unlike in the local universe, not all the dust or the cold gas is accounted for, making it more challenging. SPICA and its factor 10 to 100 in sensitivity will allow to image the faint extended cold gas/dusty disks in BCDGs in addition to detect faint C and O lines only marginally or not at all detected by Herschel.
\end{abstract}

\section{Two BCDGs out of norms}

Blue Compact Dwarf Galaxies (BCDGs) are objects considered both as local analogs to high redshift early forming galaxies and as, paradoxically, left-overs from the cosmological history. Low metallicity, low dust content, relatively high star formation rates (wrt to total mass), large gas-to-star ratios, dwarf galaxies have been observed over two decades at all wavelengths to understand their evolution, how they survived the various cosmological events that punctuated the cosmological history. Recently combining high sensitivity and spatial resolution Herschel observations (covering at the wavelength range between 50$\mu$m and 500$\mu$m) with previous MIPS observations, it was shown that the dust properties as function of the star-formation rate and metallicity in low metallicity dwarf galaxies (12+log(O/H) $<$ 8.1) behave very differently from higher metallicity dwarf galaxies and other normal galaxies.

\begin{figure}[!ht]
\begin{center}
\resizebox{0.5\hsize}{!}{
\plottwo{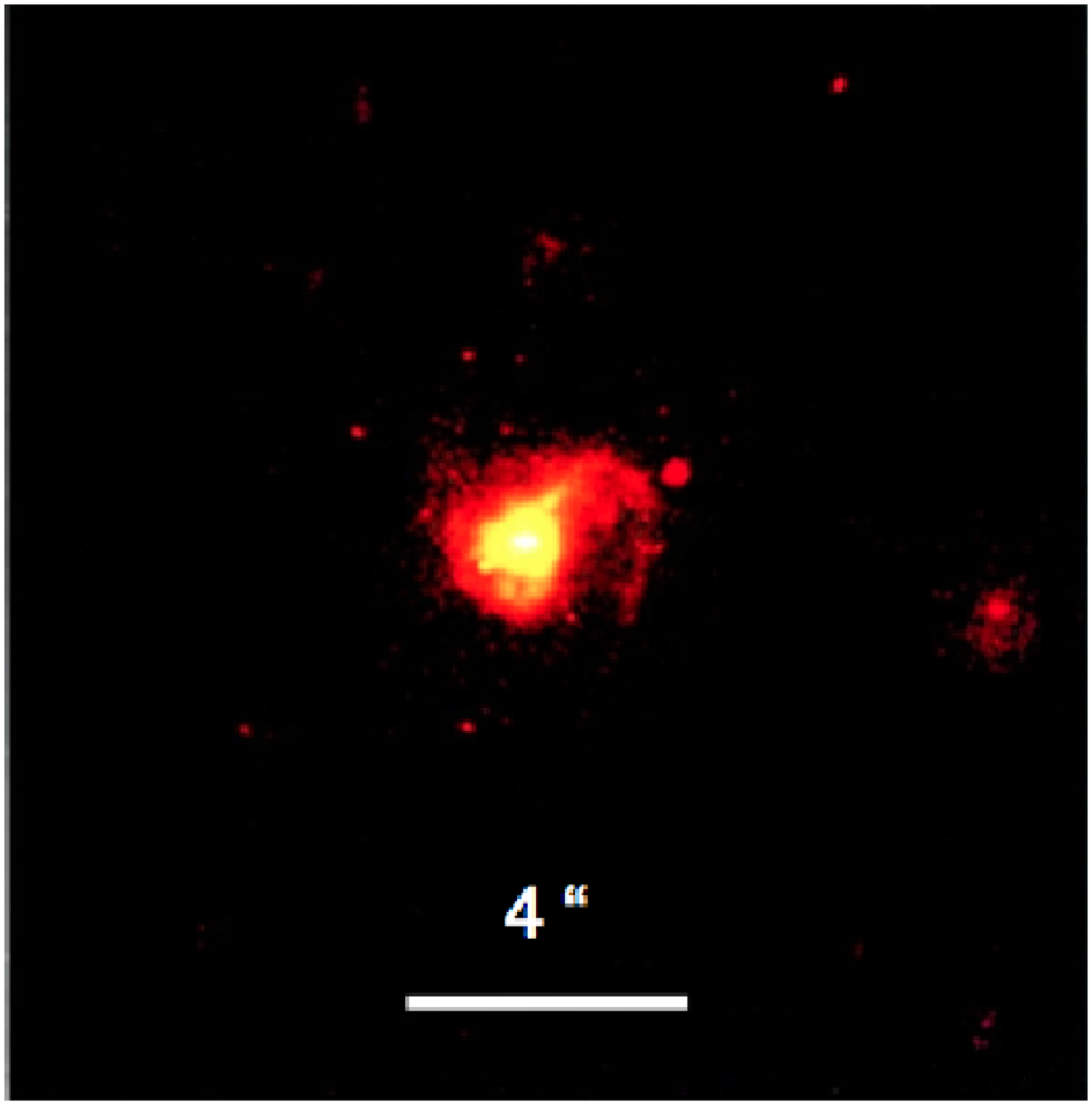}{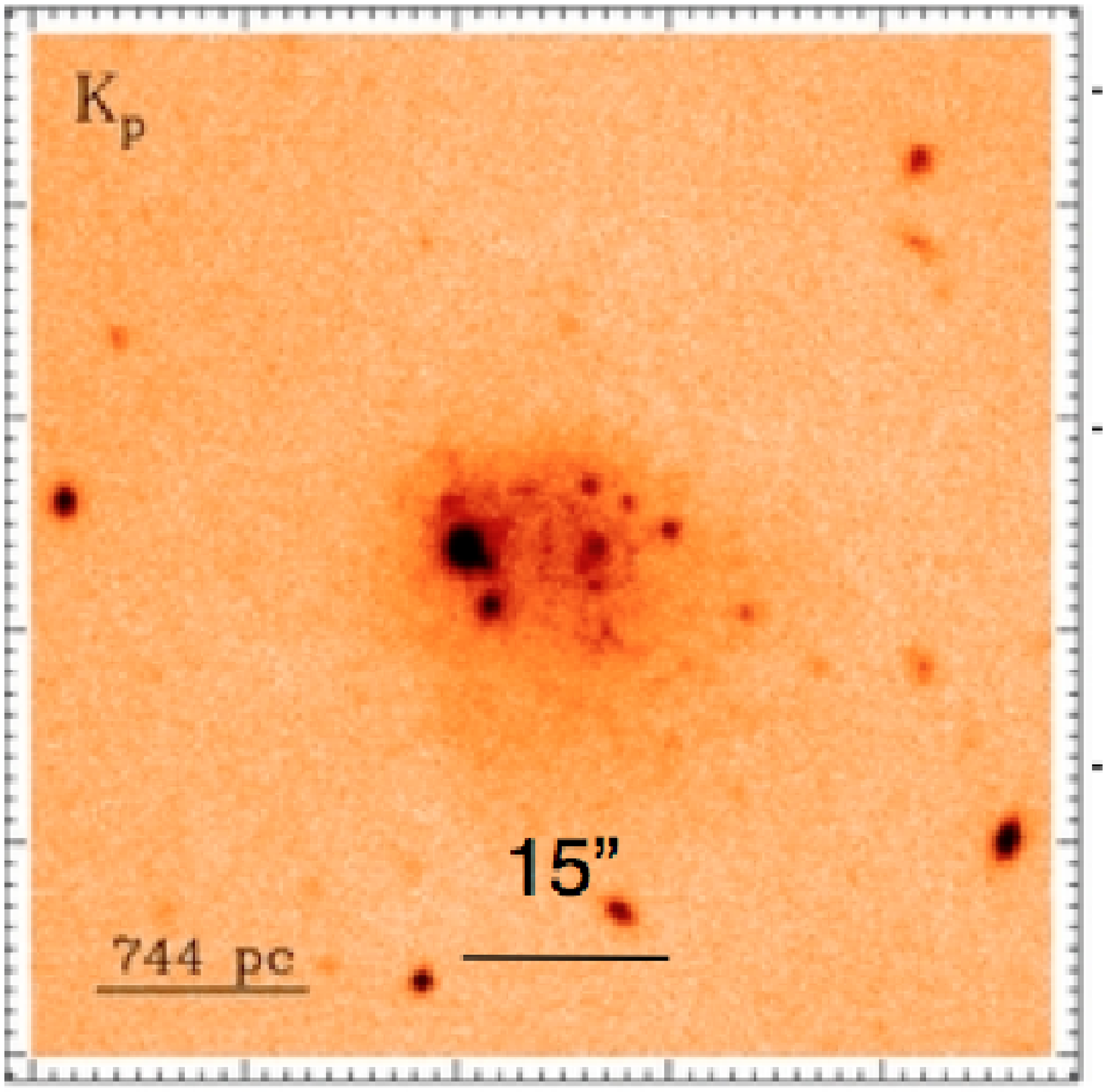}
}
\end{center}
\caption{
\textit{Left:}\, POX~186: HST VRI  
\textit{Right:}\, UM~461: K band }
	\label{fig:BCDGs}
\end{figure}

UM~461 and POX~186 are two different BCDGs with almost opposite properties: UM~461 is a low metallicity gas rich with an underlying old stellar population ($<$ 5 Gyrs) dwarf galaxy (figure~\ref{fig:BCDGs}\,\textit{Right}, \citep{lagosetal2011}, \citep{doublieretal01}) while POX~186 is a low metallicity gas poor with no apparent underlying old stellar population ultra compact dwarf galaxy (figure~\ref{fig:BCDGs}\,\textit{Left}, \citep{corbinetal2002}), \citep{doublieretal00}).  
 
\section{Dust properties in POX~186 and UM~461}

\begin{figure}
\begin{center}
\resizebox{0.6\hsize}{!}{
\plottwo{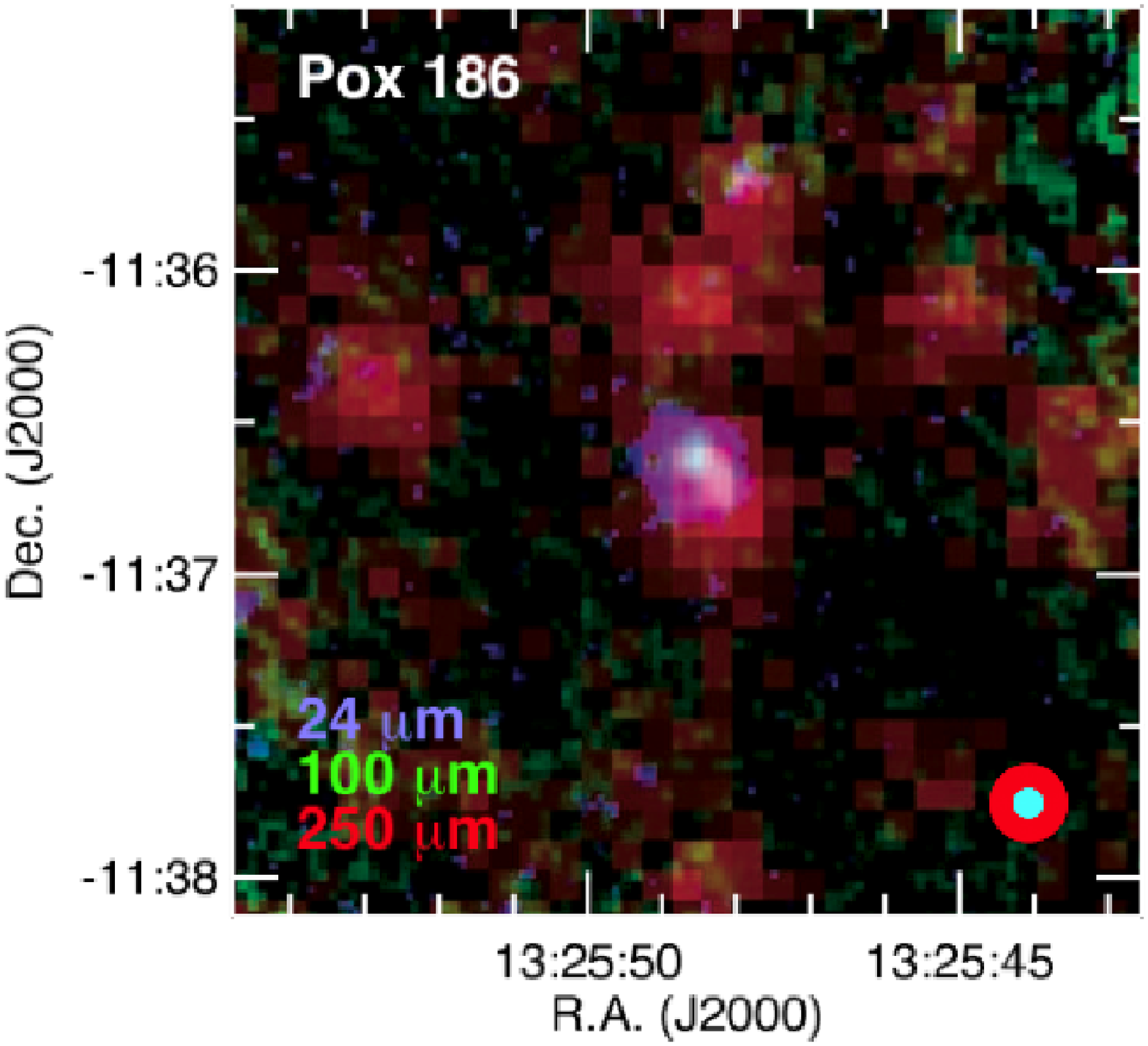}{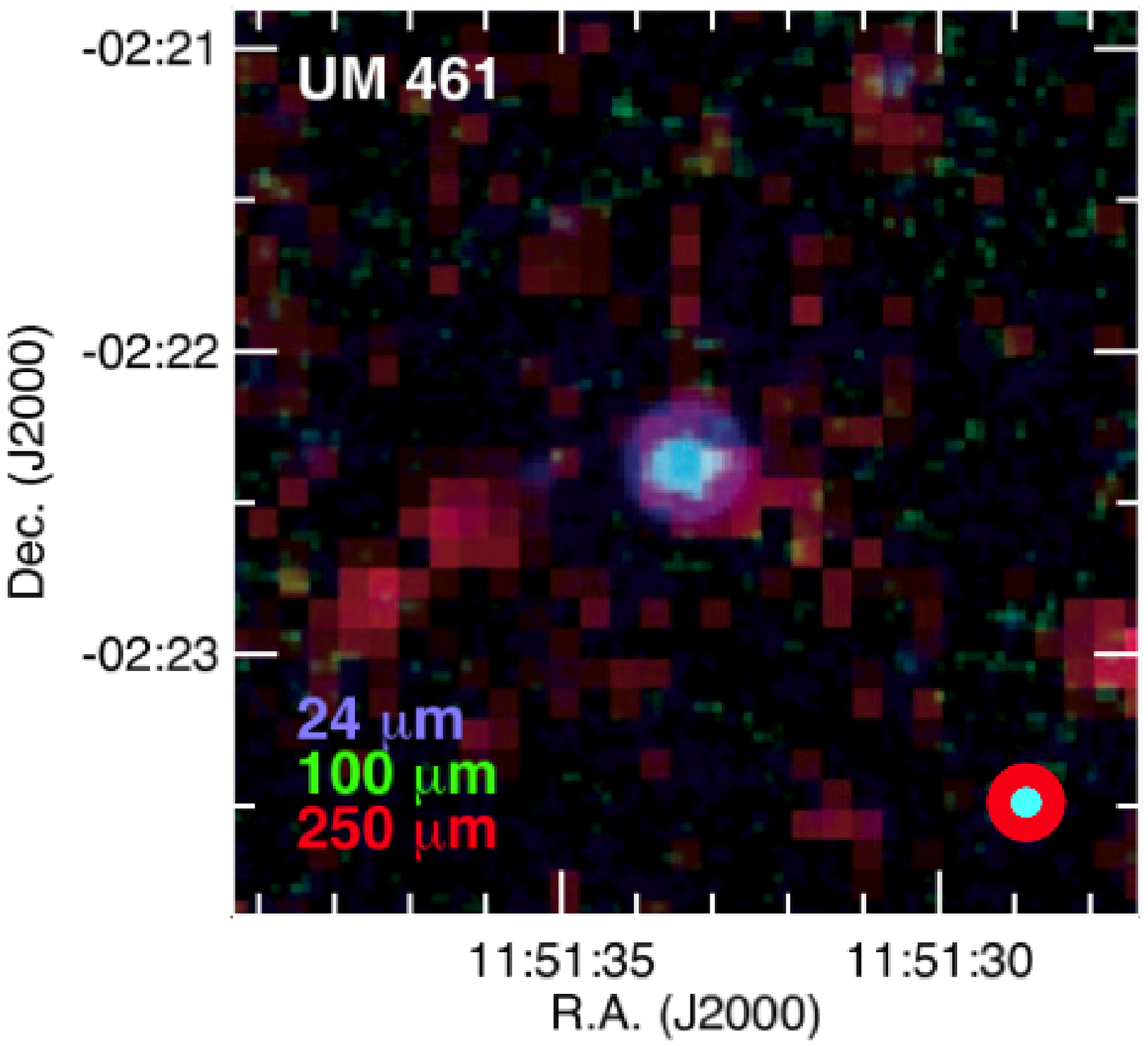}
}
\end{center}
\caption{  RGB composite image: 24$\mu$m MIPS and HERSCHEL Pacs 100$\mu$m +Spire 250$\mu$m 
\textit{Left:}\,POX~186
\textit{Right:}\, UM~461
}
	\label{fig:bcdg}
\end{figure}

POX~186 and UM~461 (figure~\ref{fig:bcdg}) were observed as part of the "Dwarf Galaxies Survey" (PI: S. Madden, \citep{maddenetal2013}). For both objects the following data were obtained using the Herschel observatory with PACS: 70$\mu$m, 100$\mu$m and 160$\mu$m and SPIRE 250$\mu$m, 350$\mu$m and 500$\mu$m.  The details of the data processing, photometry and spectral energy distribution fit is given in \citep{remyetal2013a} and R\'emy-Ruyer et al. (2013 in prep). The MIPS data at 24$\mu$m were taken from \citep*{bendoetal2012b}.

\subsection{SED}

In the MIR + FIR, POX~186 does not show the same features as in the visible (Figure ~\ref{fig:bcdg}, \textit{Left}). The galaxy displays a cold tail which origin is unclear: cold gas/dust tail/arm. POX~186 is an archetypical "dark galaxy".

\begin{figure}
\begin{center}
\resizebox{0.71\hsize}{!}{
\plottwo{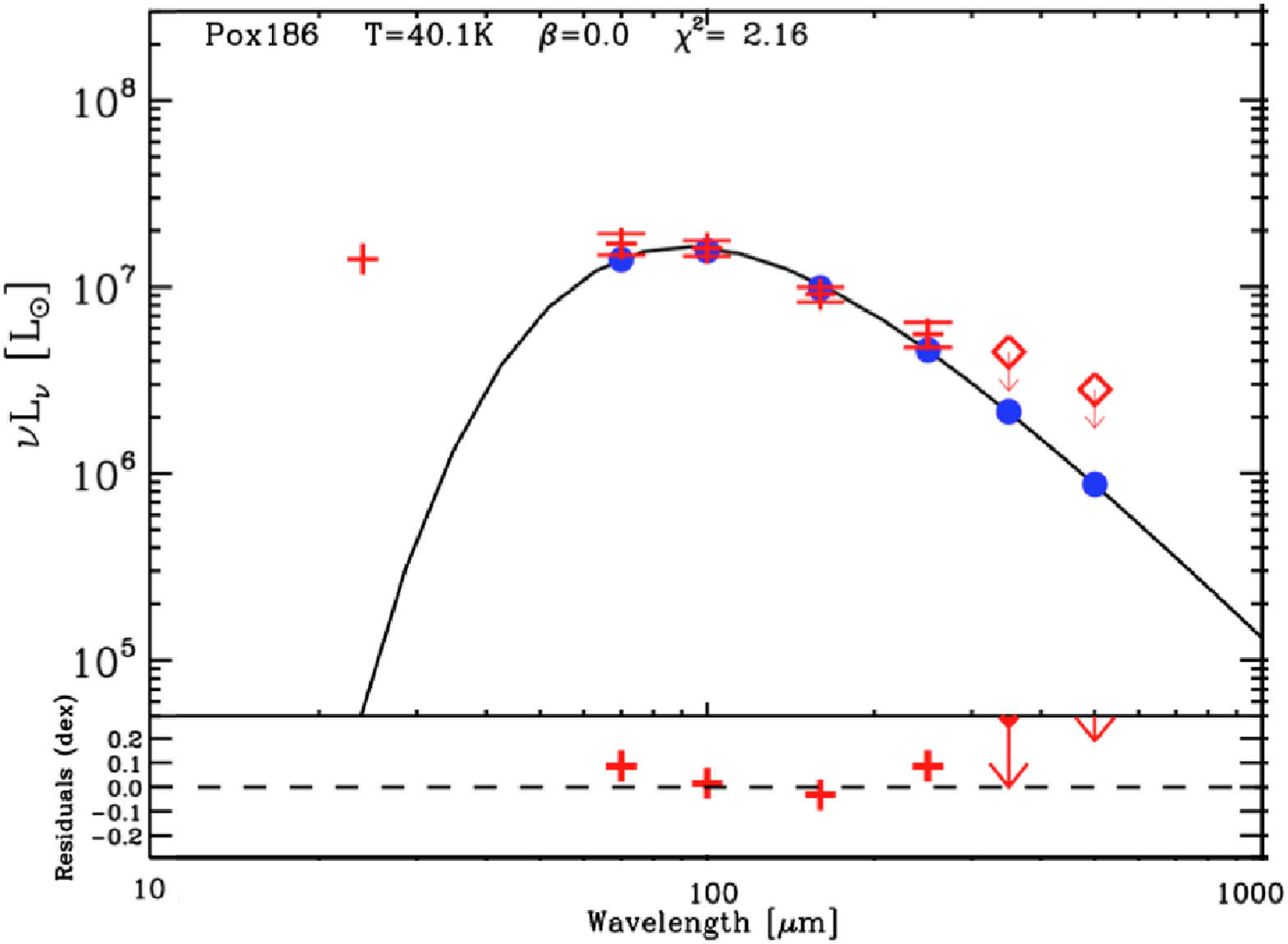}{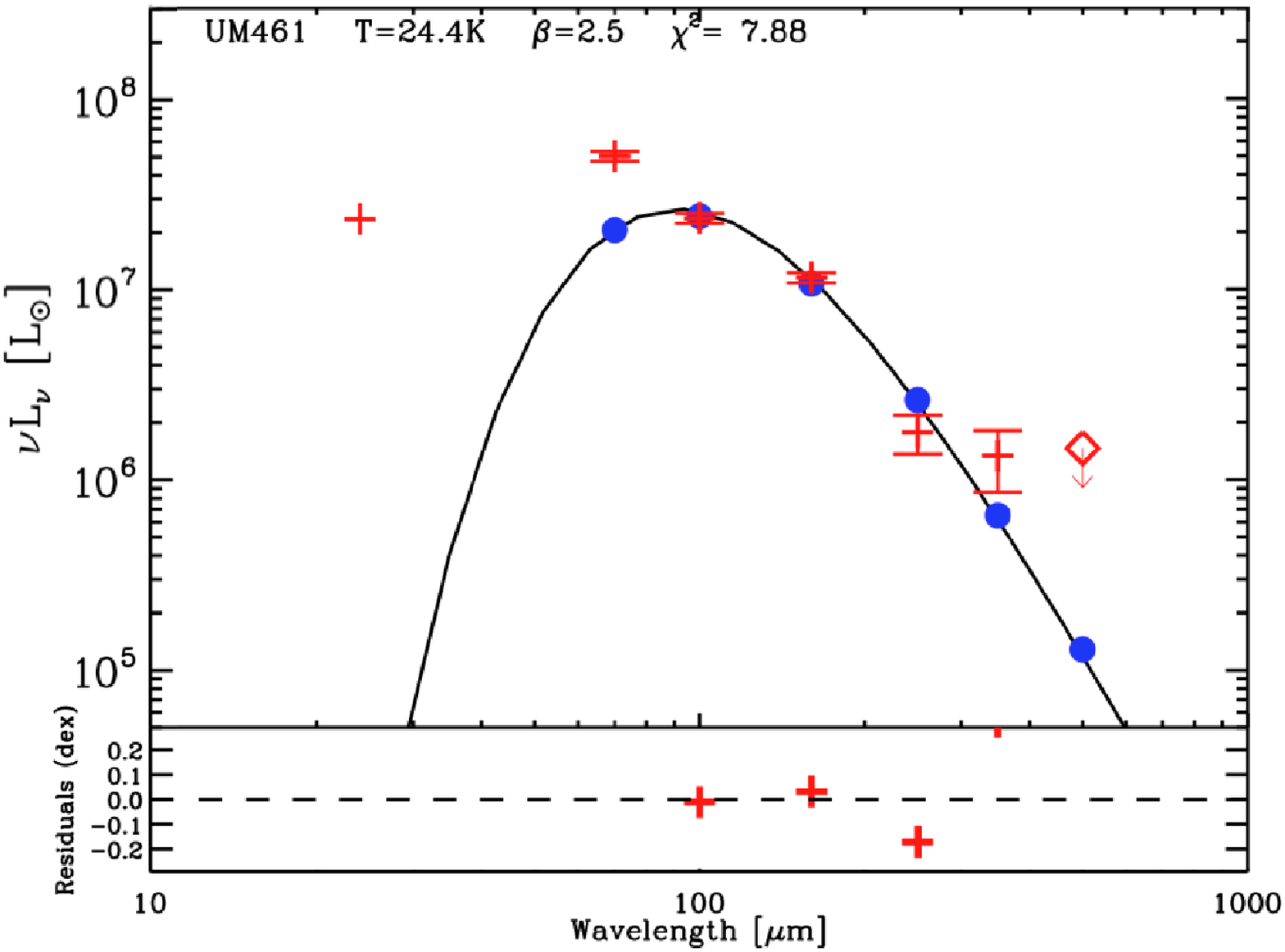}
}
\end{center}
\caption{SEDs: Observed SED (red symbols), Fitted SED 1-component  (blue symbols and black line), 2-component (yellow line)
\textit{Left:}\, POX~186
\textit{Right}\, UM~461
}
\label{fig:UM}
\end{figure}

POX~186 and UM~461 are is plotted on figure~\ref{fig:UM}. Only upper limits could be obtained for POX~186 at 350$\mu$m and 500$\mu$m.  This sets a lower limit for the dust and/or cold gas component temperatures. The blue symbols represent the fitted SED under the following assumptions: single dust component described by a modified Black-Body emission (see R\'emy-Ruyer et al. (2013 in prep) for all details). The black line represents the modified BB models for the given dust mass, dust temperature and dust emissivity.

For POX~186, the observed SED appears to  depart  from the model at long wavelengths. The flat observed SED is indicative of very different components influencing the energy distribution in the MIR and FIR. As seen in the two images: visible (figure~\ref{fig:BCDGs}\, \textit{Left}) and MIR+FIR (Figure ~\ref{fig:bcdg}, \textit{Left}) there seems to be a warm component corresponding to the central star forming region, and much colder and extended underlying disk of gas/dust. However, the integrated photometry includes both regions and higher revolution with higher sensitivity would allow the disentangle the various components. 

UM~461 shows two distinct components in the visible and near infrared which properties are different such as stellar population, star formation age and history. The Eastern star formation region shows much "younger" characteristics than the West star forming region (figure~\ref{fig:bcdg}, \textit{Right}). The distinctions are also reflected in the MIR+FIR range (figure~\ref{fig:UM}, \textit{Left}). One region (West) is much colder than the other (East).

The observed SED cannot be reproduced with a single dust component as shown in figure~\ref{fig:UM}, \textit{Right}. The fit diverges at both end of the energy distribution. 

\begin{figure}
\begin{center}
\resizebox{0.4\hsize}{!}{
\includegraphics*{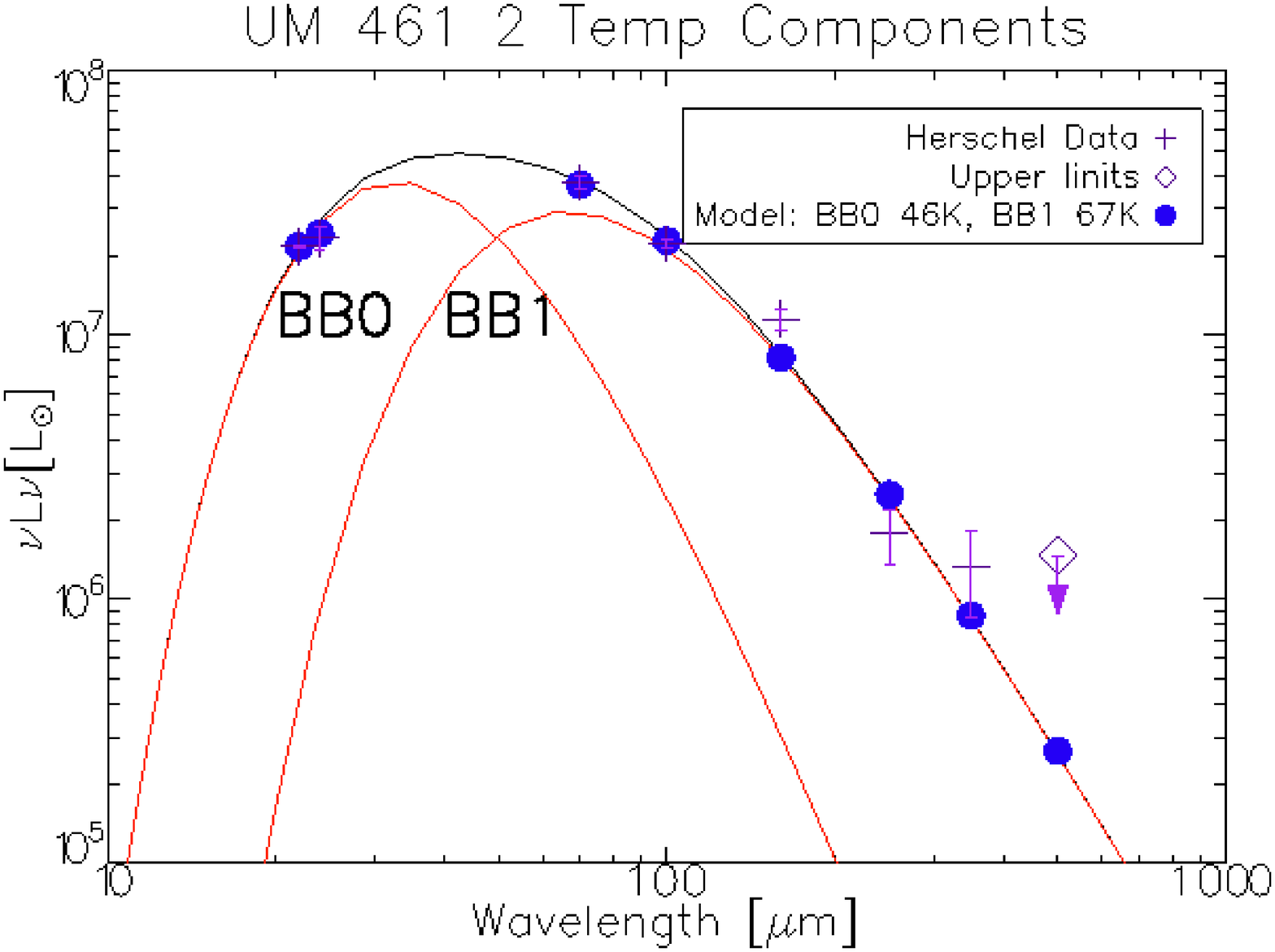}
}
\end{center}
\caption{UM~461: 2-components modified BB SED}
\label{fig:UM2}
\end{figure}

A second fit using a two-component modified black-body model was used and is shown in figure~\ref{fig:UM2}.

Contrary to POX~186, UM~461 shows the indication of hosting a warmer dust component possibly associated to the younger star formation located East. UM~461 could be a local counterpart of the high redshift SINSs galaxies inside which self- gravitating star forming regions temporarily form, then merge back into the galaxy disk.

\subsection{Anomalous objects}

\begin{figure}
\begin{center}
\resizebox{0.91\hsize}{!}{
\plottwo{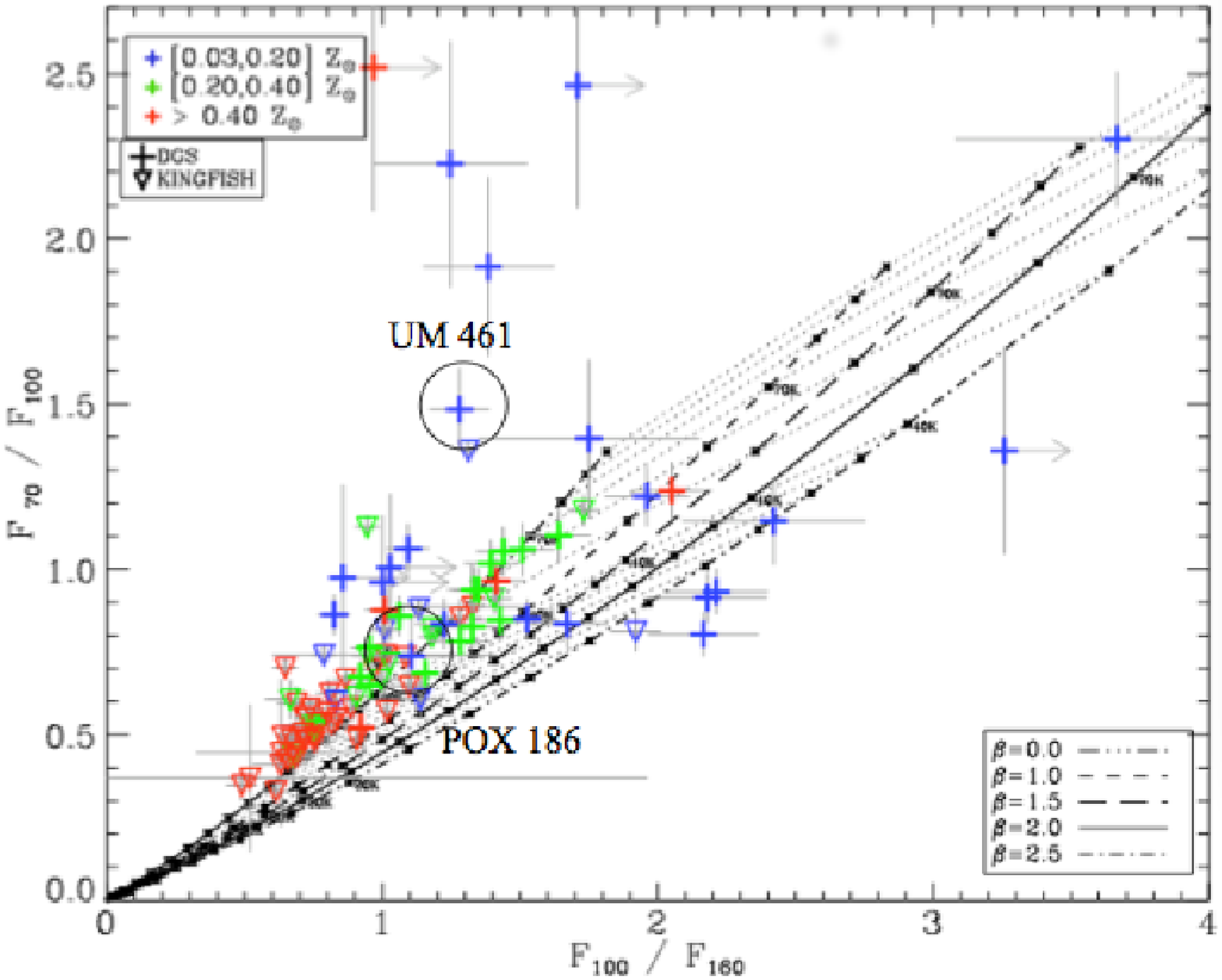}{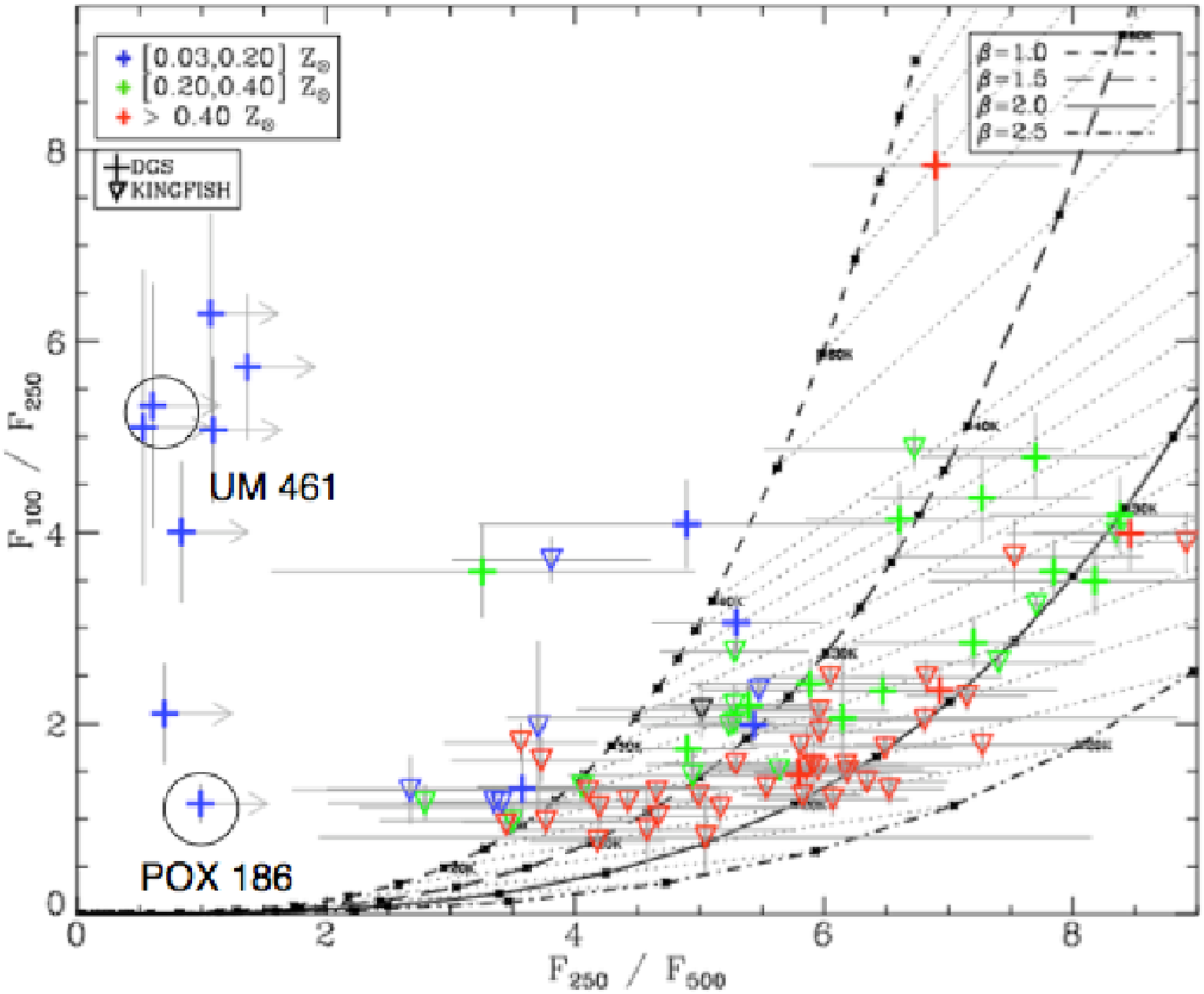}
}
\end{center}
\caption{ 
\textit{Left}\, PACS/PACS diagram : F70/F100 versus F100/F160. the colours corresponds to different metallicity bins. Crosses and downward triangles represent DGS and KINGFISH galaxies, respectively.
\textit{Right}\,  Colour-colour diagram : PACS/SPIRE diagram : F100 /F250 versus F250 /F500 .
}
\label{fig:anom}
\end{figure}

In figure~\ref{fig:anom}, the curves give theoretical Herschel flux ratios for simulated modified black bodies for $\beta$=[0.0, 2.5] and temperatures T=[0,40]K in 2K bins and T=[40,100]K in 10K bins (black dots). Lines of constant temperature are indicated as dotted lines (see R\'emy-Ruyer et al. (2013 in prep)).

Note that POX~168 and UM~461 amongst the most metal poor galaxies (from 0.03 to 0.20 Z$\odot$) are very faint and even not detected anymore at longer wavelength. However, when detected their MIR properties differ from the bulk of dwarf galaxies. The dust composition, temperature and spatial distribution may be very different due to the very low metallicity and the particulars of the star formation event.

\section{SAFARI}

 In coordination with interferometric submm-radio telescopes ( Alma (in the southern hemisphere),  Plateau de Bures) to cover the SED colder components at high spatial resolution, SPICA represents the perfect follow up tool for local BCDGs, and to search and study higher redshift (z$\sim$0.5 - 1) dwarfs.
It will allow to spatially resolve both galaxies and structures and detect down to 210$\mu$m all components responsible for the anomalous SED shape and allow multiple components fitting

\acknowledgements 
V. Doublier Pritchard wishes to thanks the SAG-2 consortium for allowing the use of GTO data, and A. R{\'e}my-Ruyer for providing the SED and SED-fits of the 2 cases here discussed.
  
\bibliography{doublier_bib}

\end{document}